# The Generation and Analysis of Tritium-substituted Methane

D. Díaz Barrero,[a,*]  T. L. Le,[b]  S. Niemes,[b]  S. Welte,[b]  M. Schlösser,[b,*]  B. Bornschein,[b] and H. H. Telle[a]

[a] *Departamento de Química Física Aplicada, Universidad Autónoma de Madrid, 28049 Madrid, Spain*

[b] *Tritium Laboratory Karlsruhe (TLK), Institute for Astroparticle Physics (IAP), Karlsruhe Institute of Technology (KIT), 76021 Karlsruhe, Germany*

[*] *Corresponding authors:* deseada.diaz@uam.es , magnus.schloesser@kit.edu

**Abstract**: *An unavoidable category of molecular species in large-scale tritium applications, such as nuclear fusion, are tritium-substituted hydrocarbons; these form by radiochemical reactions in the presence of (circulating) tritium and carbon (mainly from the steel of vessels and tubing). Tritium-substituted methane species, $CQ_4$ (with Q = H , D , T), are often the precursor for higher-order reaction chains, and thus are of particular interest. Here we describe the controlled production of $CQ_4$ carried out in the CAPER facility of the Tritium Laboratory Karlsruhe (TLK), exploiting catalytic reactions and species-enrichment via the CAPER-integral permeator. $CQ_4$ was generated in substantial quantity (>1000 $cm^3$ at ~850 mbar, with $CQ_4$ - content of up to ~20 %). These samples were analyzed using laser Raman and mass spectrometry, to determine the relative isotopologue composition and to trace the generation of tritiated chain-hydrocarbons.*

## I. INTRODUCTION

Hydrogen chemistry is very versatile and varied, and hydrogen ($^1H \equiv H$) can form molecular compounds with almost all elements in the periodic table. These are encountered in the form of ionic-bond hydrides, but also as covalent-bond compounds, like e.g. hydrocarbon molecules, water, ammonia, etc., and in multiple metallic interactions. It also participates in adsorption and absorption processes, occupying interstitial voids in crystalline and metallic networks. This wealth of interactions is due to its small size and its electronic structure; note that the chemistry of its isotopes – deuterium ($^2H \equiv D$) and tritium ($^3H \equiv T$) – is equivalent, in principle. However, for tritium (the radioactive isotope of hydrogen) one must also consider the radiochemical reactions that occur as a consequence of its β-decay

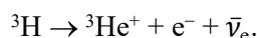

$$^3H \rightarrow {^3He^+} + e^- + \bar{\nu}_e.$$

The highly energetic decay products, i.e. the helium ions and the β-electrons, lead to the generation of a wealth of secondary ions and radicals, through impact interaction with any surrounding atoms / molecules.[1] This peculiarity of tritium modifies its chemical equilibria and reaction products with respect to the two non-radioactive isotopes (H and D), as well as the chemical kinetics of the reactions in which it participates.[2] Because of this, the products of these reactions and their concentrations are difficult to predict, in general.

   In large-scale tritium applications, such as the KATRIN (**Ka**rlsruhe **Tri**tium **N**eutrino) experiment,[3] and the future nuclear fusion experiment ITER (International Thermonuclear Experimental Reactor) – as well as its successor experiments – non-negligible amounts of short-chain tritiated hydrocarbons are generated. By and large, they contain between one and five C-atoms, which have been detected in circulating $Q_2$ process gases (with Q ∈ [T , D , H]). This occurs despite the fact that these systems have gas-filtering units – such as permeation cells (or 'permeators') – to remove unwanted non-$Q_2$ contaminants. The carbon source for this type of reaction is C-extraction from the steel pipes / vessel walls, with the reactions with tritium taking place on the steel surfaces; note that it is rather unlikely that carbon originates from leaked atmospheric $CO_2$ or CO after passing through the permeator. Contaminating molecules need to be removed from the process gases, because they may cause operational problems in large installations, being deposited in inappropriate places; or they may affect other processes, such as e.g. the formation of plasmas.[4,5] In order to set up means for the removal / elimination of hydrocarbon contaminants, it is necessary to study their chemistry, and determine their molecular structure, as well as the reactions that these molecules undergo among themselves, with the surfaces and other parts of the systems exposed to them.

A good starting point for any *in situ* / inline study of the tritium – carbon chemistry is the generation of tritium-substituted hydrocarbons in sufficient quantity, to allow one to perform precision spectroscopy on them and thus obtain relevant analytical knowledge. The tritium-substituted methane species, $CQ_4$ (with $Q \in [T, D, H]$) are of particular interest since, in general, they constitute the dominant species in tritium-circulation systems.[5] It should also be noted that $CQ_4$ may serve as precursors for forming chain-hydrocarbon molecules.

Despite being studied for quite some time, detailed spectroscopic knowledge about tritiated methane molecules is still sparse; this is because the available information corresponds to very low concentrations and very small sample quantities (see Refs. 1, 6, 7). In those experiments, mixtures of methane and tritium were left to react for up to a few days; the final products were analyzed by mass spectrometry and Raman spectroscopy.

In order to overcome the problem of only minute amounts of tritiated $CQ_4$, we utilized their production – in sufficient quantity for high-resolution laser Raman spectroscopy – in the CAPER-facility (CAprice PERmcat)[8] at the Tritium Laboratory of Karlsruhe (TLK). Quantities of >1000 cm$^3$ at ~850 mbar, with $CQ_4$-content of up to the order of ~20% (conservative estimate), were produced under controlled conditions. In this paper we describe the production procedure and the detailed analysis of samples, using mass spectrometry and Raman spectroscopy.

## II. EXPERIMENTAL

### II.A. System fundamentals

The aim of this experiment was to produce $CQ_4$ as pure as possible, and in sufficient quantity for precision analysis. Based on the aforementioned experiments we knew that the production of a chemically pure compound was not possible when relying on the hypothetical isotope exchange reaction in a simple binary mixture of gases, i.e.

$$CH_4 + 2T_2 \leftrightarrow CT_4 + 2H_2 \qquad (1)$$

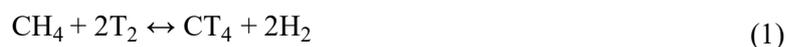

The reactions that actually take place are chemical equilibria that result in a set of products in which the H-atoms in methane are gradually replaced by T-atoms. A quasi-stable state is reached when the concentrations remain relatively constant over long periods of time, mainly consisting of the chemical family of $CQ_4$, in our case $Q \in [T, H]$:

$$CH_4 + 2T_2 \leftrightarrow CT_xH_{(4-x)} + 2Q_2 \quad \text{(with } x \in [1, 2, 3, 4]\text{)}. \qquad (2)$$

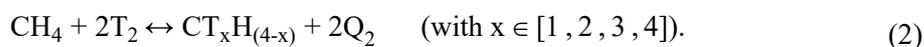

Unfortunately, in the presence of tritium, equilibria are not stable but continue to evolve over time due to the tritium activity. If any mixture is stored for a long time, e.g., 18 months in the case of the experiments carried out by McConville and Menke,[1] the observation is that the tritium and carbon concentrations in the gas phase diminish and "deposits" on the surfaces of the sample containers occur. In fact, one could argue that hydrocarbons "polymerize" at the surface. Indeed, such deposits were observed in our experiments, specifically on the windows of the Raman cell used in the analysis setup, although they appeared over much shorter time scales than those reported by McConville and Menke.[1] Note, however, that little substantiated information is available on the nature of such deposits.

Unlike in previous experimental work, we deliberately altered the chemical equilibria, to shift and direct them towards the product of highest interest, namely $CT_4$. As such, our procedure is completely different to that utilized in earlier experiments: instead of relying on "passive" exchange reactions and associated equilibration, we used the CAPER facility to perform "catalysis-driven" synthesis.

Here it is worthwhile to recall some construction and operational features of CAPER; for details see e.g. Ref. 8. CAPER was conceived as the experimental verification of the processing of the Tokamak exhaust processing (TEP) concept for ITER, whose absolute requirement was that the maximum tritium concentration in the flue gas is less than ~$3.7\times10^{10}$ Bq/m$^3$. Note that in terms of tritium concentration this corresponds to the reduction of the tritium concentration by a factor of ~$10^6$.

CAPER achieves this in three steps, in batch operation. In the first step, the effluent gases pass through a Pd/Ag membrane that is 100% selective for $Q_2$ in the permeator, where $Q_2$ (with $Q \in [T, D, H]$) is separated from molecular impurities. Two effluent streams are generated, one containing nearly all of the $Q_2$ (for later enrichment of tritium), and one in which the non-$Q_2$ contaminants are removed. In the second step, these contaminants are processed in a closed loop, using a combination of heterogeneous catalysis and permeation, to extract the $Q_2$ resulting from the catalysis. The third step is carried out in the PERMCAT reactor,[9] which simultaneously combines the permeation through the Pd/Ag membrane and the catalytic bed in a so-called "isotopic swamping" process, as used in tritium processing for the removal of impurities from the process gas.[9,10]

The use of the CAPER facility allowed us to incorporate heterogeneous catalysis into the synthesis of $CQ_4$, through the isotope exchange reactor containing a suitable catalyst. A second important feature of CAPER is that it provides the capability to regulate the flow rates of reagents fed into the reactor. In addition, the permeator provides the possibility of selectively eliminating products of the isotopic exchange reaction, namely $Q_2$. In this way one is able to move the reaction equilibrium in the direction of interest, i.e. ultimately increasing the concentration of $CQ_4$ in the final mixture.

Amongst other interesting features, catalysts have the ability to direct the reaction towards certain products as well as to improve the reaction kinetics, thus significantly reducing reaction times. The catalyst used in the study described here is a commercial nickel catalyst on Kieselguhr (a porous diatomaceous-earth substrate used as a support for catalysts, to maximize a catalyst's surface area and activity), with an approximate nickel content of 50%. In general, this type of catalyst is used for methane cracking.

The proposed mechanism of action of the aforementioned type of catalyst is the formation of carbon filaments on the active centers of the catalyst.[11]

As methane comes into contact with the catalyst, carbon chains grow, keeping nickel at the tip, which sustains the catalysis. The diffusion of carbon on the surface of the catalyst generates carbon deposits that end up deactivating its catalytic action. The operational recovery of this type of catalysts is carried out using hydrogen, among other substances. Hydrogen adsorbs reversibly on nickel, occupying active centers and displacing carbon. It competes with it for the active centers, and decreases the efficiency of the cracking reaction, and thus carbon deposits. Exploiting these two processes, the chemical environment of the catalyst can be modified, to either activate or reduce it. Note that, in our case, we use tritium instead of protium.

Hence, if the catalyst that has been previously saturated with $T_2$, $CH_4$ will be retained at its surface; therefore, the isotope exchange reactions will be favored instead of the cracking reactions,[10] thus yielding $CT_xH_{(4-x)}$.

## II.B. Sample preparation

Before starting each step of the synthesis, the catalyst has to be activated and reduced so that it is in optimal condition to produce the exchange reaction. For this, a stream of high-purity $T_2$ (>97 %) is passed through the system for at least 18 h, at a temperature of 375-400 °C. In order to try minimizing the presence of longer-chain hydrocarbons in our samples, the synthesis process proceeds in several stages, as can be seen in the three parts of Figure 1. The key elements involved in the synthesis are (i) two vessels for high-purity tritium and for the collected reaction mixture, respectively; (ii) the permeator; and (iii) the exchange reactor. Note that the valves highlighted in green are open and allow the passage of gases, and those highlighted in white are closed and thus block passage. Note also that, the valve marked SCV constitutes a unidirectional stop-check valve assembly with backflow protection, to prevent contamination of the $T_2$-vessel.

The reaction mixture obtained in each step is retained, as it becomes the starting reagent for the following step. Note that here we refer to the last synthesis run in a series of optimizing trials; the data shown in this publication are from this specific attempt. The relevant CAPER configurations for the sub-step producing Samples 1, 2 and 3 are shown in the top, center and bottom parts of Figure 1, respectively.

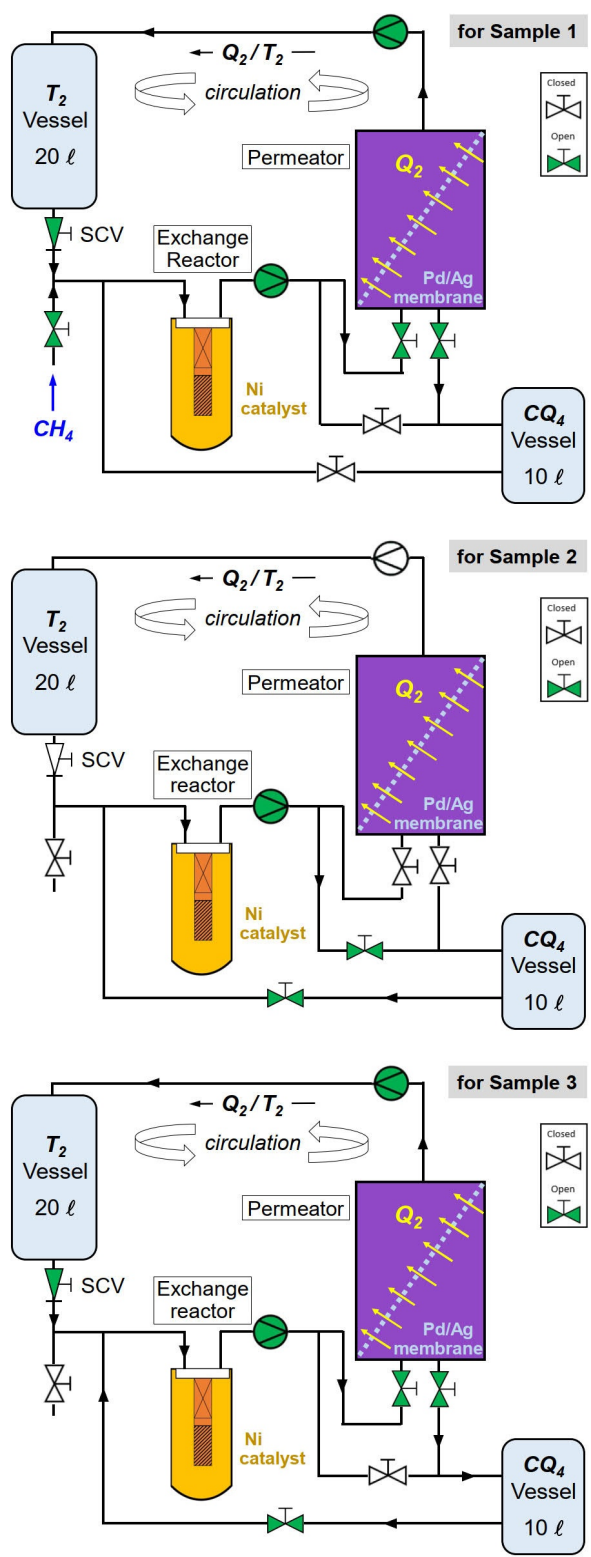

**Fig. 1**. Simplified schematic diagram of the CAPER sub-sections used in the production / enrichment of $CQ_4$ (note that under certain reaction conditions longer-chain T-hydrocarbons are generated as well). The gas circulation paths for the individual, sequential phases are indicated by arrows; they are determined by closing / opening relevant valves. SCV = unidirectional stop-check valve assembly. For further details, see text.

Accordingly, the starting Sample 1 is prepared using a flow rate of $CH_4$ of ~0.25 mol/h (of purity >99.99 %). This flow is mixed with tritium prior to entering the reactor; its flow rate was ~4 mol/h, with tritium purity >97 %. The reactor feed is maintained for ~1 h, at a temperature of 375-400 °C.

In continuous operation, the reaction products are pumped to the permeator where the compounds $Q_2 = T_2$, HT and $H_2$ are separated from the mixture. The non-$Q_2$ remainder is transferred to the $CQ_4$-vessel, being stored for the next stage of the synthesis.

Note that with the partial extraction of $Q_2$ from the mixture the exchange reaction becomes favorable for the generation of $CQ_4$ because (i) it shifts the equilibrium towards the products, avoiding the backward reaction towards re-formation of $CH_4$ (see equation 2); and (ii) it reduces the volume of the sample while increasing the fraction of $CQ_4$ in the sample. Note also that in order to keep Sample 1 at tritium saturation conditions and to favor the isotopic exchange reaction, for storage it is diluted to 50 % with high-purity $T_2$ (>97 %).

The production of Sample 2 starts with the diluted Sample 1; the relevant gas circulation path is shown in the central part of Figure 1 (note that in this procedural step the permeator is bypassed). In this second step, the gas mixture passes only through the reactor; the reaction time and temperature conditions of the reactor are kept the same as in the first step. All gases leaving the reactor are collected in the $CQ_4$ vessel, to be used in the third step of the synthesis.

The path for the production of Sample 3 is shown in the lower part of Figure 1. The mixture from the previous stage (Sample 2) is fed back into the exchange reactor and more tritium is added at a flow rate of ~6 mol/h. Circulation is maintained for ~3 h, keeping the same temperature conditions as before. Note that now, with the permeator in the loop once more, the reaction products are separated into two effluents again, i.e. $Q_2$ and $CQ_4$ are directed towards their respective vessels for storage.

Overall, about 0.21 mol of product gas mixture was generated, which is equivalent to ~$5 \times 10^3$ cm$^3$ for standard pressure and temperature conditions. This was more than enough to conduct an extensive series of analytical measurements, to characterize the gas composition and to record high-resolution Raman spectra.

**II.C. Measurement protocol**

At each step of the synthesis, aliquots of the three produced samples were transferred to tritium-compatible Raman cells and sample cylinders; these were transferred to the TRIHYDE facility of the TLK for analysis. Note that TRIHYDE is a tritium-compatible system designed for gas mixing and analysis. It incorporates a range of measurement system devices, including high-resolution laser Raman spectroscopy and mass spectrometry.[12] The samples were measured immediately after they were prepared and were not stored; thus, they should constitute a reliable snapshot of the concentrations and majority species at each stage of the synthesis.

First, the Raman spectra are measured, using the TRIHYDE-incorporated laser Raman (LARA) measurement instrument. This is assembled around a glovebox appendix,[13] containing the Raman sample cell. It comprises a green diode-pumped solid-state laser (GEM, *Laser Quantum*), with emission wavelength $\lambda_L = 532$ nm and continuous output power $P_L = 2$ W; Raman light collection at 90°, with fiber-bundle coupling to the Raman spectrometer (SP2150 spectrograph with Pixis400B CCD detector, *Princeton Instruments*).

All our Raman cells have the same volume of ~7 cm$^3$; they are filled to around ~850 mbar of sample gas. Note that, for the Raman spectral data shown in Figure 2, 20 spectra of 90 s accumulation time each were recorded (and averaged), in the spectral range of interest 1500-4200 cm$^{-1}$.

After completion of the Raman measurements, the sample gas is blead from the cell into the mass spectrometer (quadrupole residual gas analyzer HPQ3S, *MKS Instruments Inc*) via inlet dosing (UDV 040, *Pfeiffer*); the measurement pressure was adjusted to about 10$^{-4}$ mbar. In the instrument range 1-100 amu, 50 individual spectra – of 30 s accumulation each – were recorded (and averaged); these mass spectral data are discussed later in Section III.B.

## III. RESULTS

### III.A. Raman spectroscopy

The sequence of spectra from the start of the catalytic $CH_4 + T_2$ synthesis (Sample 1) to its end (Sample 3) – shown in Figure 2 – reveals three distinct series of features; these we discuss separately.

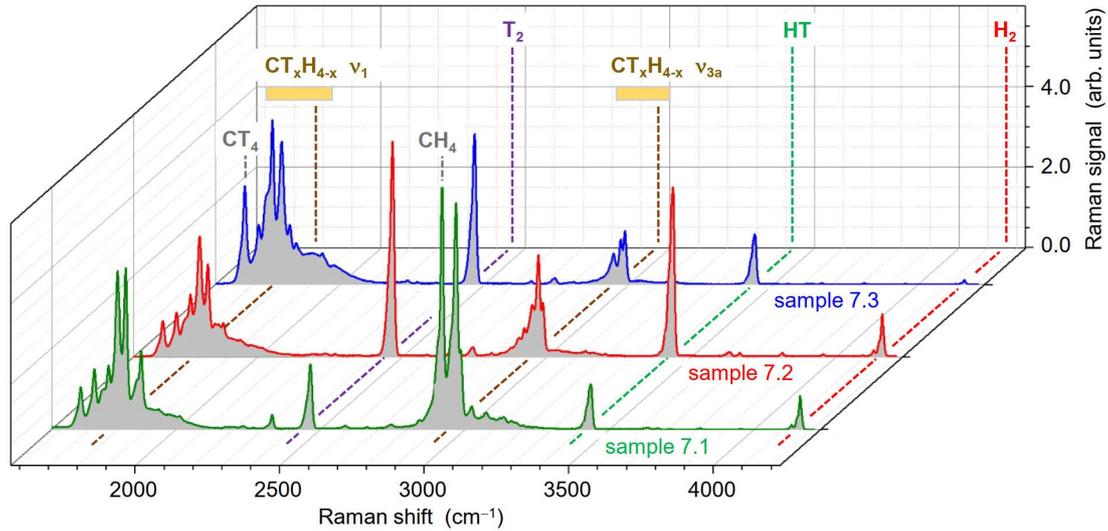

**Fig. 2**. Data of analysis of the Raman spectra for the sequence Sample 1, 2 and 3, recorded after the individual $CQ_4$-enrichment steps; for details, see text.

First, all three spectra contain the characteristic $Q_1$-branch lines of the molecular hydrogen isotopologues, as expected for this type of synthesis, i.e., from $T_2$ (at ~2466 cm$^{-1}$), HT (at ~3435 cm$^{-1}$) and $H_2$ (at ~4159 cm$^{-1}$). Their amplitudes differ significantly, indicating the evolution of concentrations for these molecular constituents. While $T_2$ is present in the starting mixture by default, HT and $H_2$ are products generated during the synthesis reactions. The source for the generation of these hydrogen isotopologues is $CH_4$, in which H-atoms are substituted by T-atoms in the course of isotope exchange reactions. Thus, in the Raman spectrum corresponding to the initial Sample 1, also the Raman peak corresponding to the $\nu_1$-band of $CH_4$ can be observed (at ~2919 cm$^{-1}$). This peak diminishes as the synthesis progresses; already in the spectrum for Sample 2 it has almost completely disappeared, with $CH_4$ increasingly being converted into $CT_xH_{(4-x)}$. Note that the spectral bands displaced towards larger Raman shift, relative to $CH_4$ $\nu_1$: these are associated with the $\nu_{3a}$-bands of those tritium-substituted methane molecules. Note also that, the highest content of $T_2$, HT and $H_2$ – with respect to the rest of the species present in the mixture – is observed for Sample 2. This is because for its preparation the permeator had been bypassed, which otherwise would have largely removed $Q_2$ from the mixture.

The set of Raman spectral features in the range 1650-1900 cm$^{-1}$ is dominated by the contribution from the symmetric, normal vibrational mode $\nu_1$ of the $CT_xH_{(4-x)}$ family. Note that the peak of the fully tritiated methane, $CT_4$, increases at the expense of the other T-methane species as the gas mixture passes through the catalyst, and the permeator removes the $Q_2$-products from the medium, thus preventing hydrogen from reverse substitution.

For further, semi-quantitative interpretation of the spectra one should recall that the Raman signal, $S_R$, can be approximated by the simplified expression

$$S_R \propto \tilde{\nu}_L \tilde{\nu}_R^3 \cdot \Phi_{if} \cdot N_i \cdot I_L \qquad (3)$$

where $\tilde{\nu}_L$ and $\tilde{\nu}_R$ are the laser and Raman transition energies, respectively; $\Phi_{if}$ is the Raman transition probability; $N_i$ is the molecular particle density; and $I_L$ is the excitation laser power. This expression (3) can be used to extract the relative molecular concentrations from the spectra, in principle. However, while for tritium-containing $Q_2$, values for the transition probabilities $\Phi_{if}$ are known,[12] those for the tritium-substituted methane species $CQ_4$ are not available; thus, quantitative interpretation is rather difficult at present.

For a better understanding of the evolution in the concentrations of the $CT_xH_{(4-x)}$ family one may zoom into the $\nu_1$-region of the Raman spectra, as shown in Figure 3.

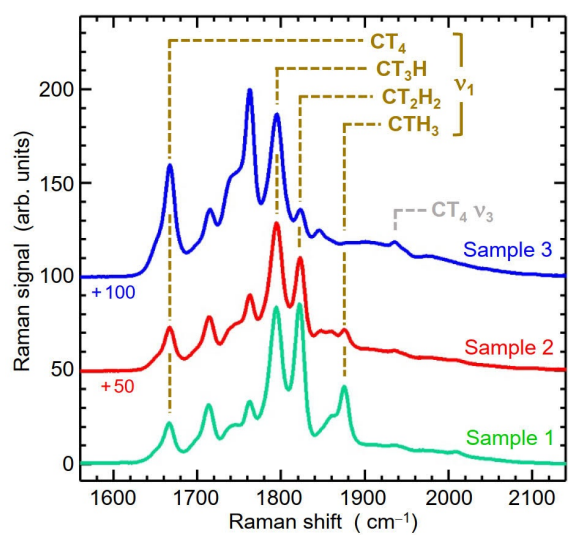

**Fig. 3**. Detail of the group of peaks corresponding to the symmetric normal mode of vibration $\nu_1$ of the $CT_xH_{(4-x)}$ family of compounds.

After the first passage of the sample gas through the catalyst reactor, the predominant species in Sample 1 are those with the highest hydrogen content, i.e. $CTH_3$ and $CT_2H_2$; but already a substantial quantity of $CT_3H$ is generated, together with smaller amounts of $CT_4$.

In the preparation of Sample 2 – for which the gas mixture passes through the catalyst for a second time but bypasses the permeator – the T-methane species, which dominated in Sample 1, start to diminish, along with the initial reagent $CH_4$. Conceptually, this is not surprising: recall that for the second step Sample 1 was diluted to 50% in high-purity $T_2$. Thus, the dominant species in the chemical environment of both the catalyst and the gas mixture is tritium. Consequently, the action of the catalyst and the balancing with surplus tritium drive the active centers towards saturation with tritium, i.e. effectively exchanging most H-atoms with T-atoms.

In the last step of the synthesis, i.e., in generating Sample 3, the relative concentration for the species of highest interest, $CT_4$, increases by almost a factor of three with respect to the previous phase. The two main reasons for this increase are:

(i) the gas fed to the reactor in this phase shows $CT_3H$ as the majority species within the $CT_xH_{(4-x)}$ family, and thus a significant fraction of exchange reactions yields $CT_4$; and

(ii) the circulation time, and thus the time available for exchange reactions, is three times longer than in the two previous phases.

The nearly linear relation between the growth in $CT_4$ concentration and the time over which exchange reactions progress invites the notion that, at least on the time scale of the experiments discussed here, it is the catalytic conversion which dominates, and that the influence of tritium decay on the reactions may be treated as being secondary.

It should be stressed that the synthesis process critically depends on the state of the catalyst. For this, optimal reaction conditions of the catalyst (activated or reduced) need to be established at the beginning of each phase. Of equal importance is that an optimal reaction environment in the medium is maintained, i.e., that fresh high-purity $T_2$ needs to be injected, and that the products HT and $H_2$, generated by the catalysis reactions, are removed. By this means, the catalyst remains in optimal conditions for longer, and the reaction favors the increase of tritiated methane molecules.

If the catalyst does not operate under optimal conditions, it would likely deactivate. This would result in mixtures containing lower concentrations of $CT_xH_{(4-x)}$ ($CT_4$ in particular), and higher amounts of hydrocarbon chain molecules; this was observed in some of our earlier synthesis attempts.

## III.B. Mass spectrometry

With regard to the molecular isotopologues $Q_2$, it can be seen that Sample 2 is the one with the highest quantity of $T_2$ (at 6 amu), HT (at 4 amu) and $H_2$ (at 2 amu), and their atomic fragments, as was the case for the Raman spectroscopic data shown in Figure 4.

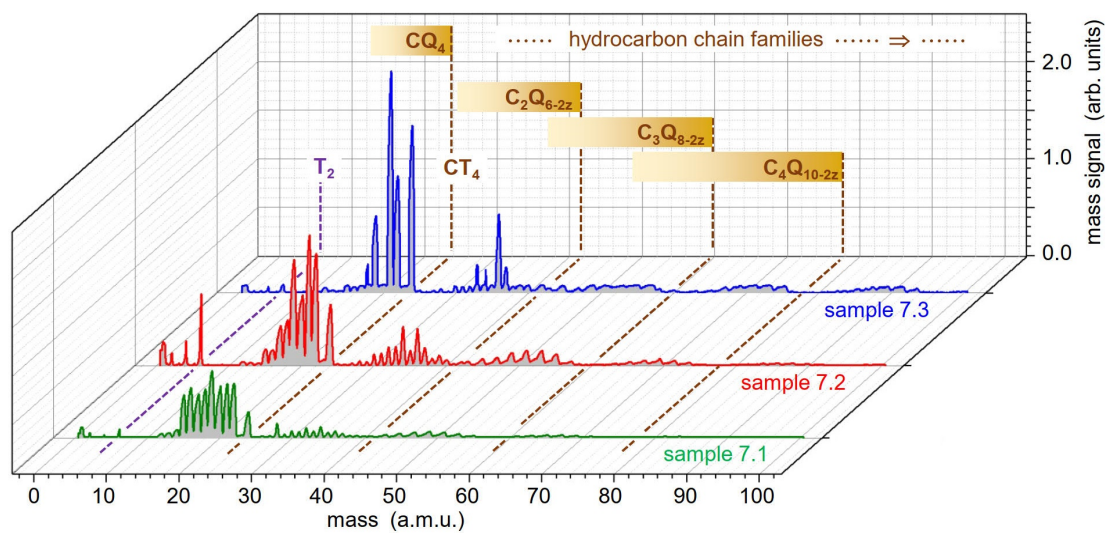

**Fig. 4**. Data of analysis of the mass spectra for the sequence Sample 1, 2 and 3, recorded after the individual $CQ_4$-enrichment steps; for details, see text.

For the family of tritium-substituted methane molecules, $CT_xH_{(4-x)}$, the mass spectra show that while all members of the family are present, (semi-) quantification becomes somewhat complex – in contrast to the Raman data. This is because of fragmentation into "daughter" molecules, in which successively Q-atoms are stripped from the "parent" molecule, leading to the same daughter molecules from different parent molecules. Overall, feature-rich mass spectra result if many, or all, $CT_xH_{(4-x)}$ parents are present in the gas mixture (for mass spectral patterns see Ref. 14).

For Sample 1, for which hydrogen is still plentiful, all $CT_xH_{(4-x)}$ parents and their daughter products are observed, giving rise to a series of incremental mass peaks ($\Delta m$ = 1 amu), ranging from atomic carbon, C (12 amu), to fully tritiated methane, $CT_4$ (24 amu). The exception is mass 23 amu, for which neither a parent nor a daughter fragment of $CT_xH_{(4-x)}$ exists. As the synthesis process progresses, the mass spectra become sparse, until for Sample 3 basically only two parents and two of their daughter fragments prevail, namely $CT_4$ (24 amu) / $CT_3$ (21 amu) and $CT_3H$ (22 amu) / $CT_2H$ (19 amu). This confirms the findings from the Raman spectra quite well.

It is worth noting that, the aforementioned filament phenomenon manifests itself in the mass spectra, too. As the overall interaction time between the catalyst and the mixture increases, so does the concentration of carbon chain molecules (number of C-atoms n>1). These can be saturated hydrocarbons, i.e., $C_nH_{(2n+2)}$ (with only C–C single bonds in the chain ≡ alkanes); or unsaturated hydrocarbons, i.e., $C_nH_{2n}$ (with one C=C double bond in the chain ≡ alkenes); or $C_nH_{(2n-2)}$ (with one C≡C triple bond in the chain ≡ alkynes). Of course, more than one unsaturated bond may be encountered, giving rise to a range of additional hydrocarbon families.

For the particular family of molecules with two C-atoms (n=2), $C_2Q_{6-2z}$ (with z indicating the number of unsaturated bonds) the following, interesting feature is observed. In the mass spectrum of Sample 3 the peak with mass 36 amu stands out, exhibiting quite a large amplitude which has evolved over the duration of the synthesis process. It can be assigned to the chemically rather stable molecule $C_2T_4$, i.e., fully-tritiated ethylene. While exact quantification was not possible, its concentration seems to be substantial in comparison to the signal amplitudes of other hydrocarbon molecules.

This leads us to think that some of the lower intensity Raman peaks, which we could not yet identify in the spectra, might actually originate from the aforementioned $C_2T_4$.

## IV. DISCUSSION

As suspected quite from the beginning, a chemically pure compound – in principle we had wished for "pure" $CT_4$ – could not been obtained. However, the multistep synthesis strategy has allowed us to shift the chemical equilibria and direct the reaction products towards the species with the highest tritium content, the majority being $CT_3H$ and $CT_4$ with almost similar magnitude (compare the amplitudes in both the Raman spectra shown in Figures 2 and 3). The concentrations of the different species can be estimated from the Raman spectral data; by and large, these were corroborated in the mass spectrometry analysis.

As stated earlier, the species concentrations can be reasonably well quantified from the (integrated) $Q_1$-branch Raman data, based on equation (3). However, for this the Raman transition probabilities $\Phi_{if}$ for each species need to be known. Unfortunately, these are not known for the tritium-substituted methane molecules. Regardless, one can make at least a crude estimate for the $CQ_4$ content in Sample 3. Using the observed relative transition probability, $\Phi_{rel}$, for the non-tritiated species $CH_4$ (with $\Phi_{rel}(CH_4) = 6.0$) and $H_2$ (with $\Phi_{rel}(H_2) = 2.4$),[15] and assuming that $\Phi_{rel}(CH_4) \approx \Phi_{rel}(CQ_4)$, one can estimate the $CQ_4$-content of Sample 3 to be ~20 %. From this estimate, the yield for this isotope exchange reaction, i.e. for the conversion of $CH_4$ to $CT_3H$ and $CT_4$ (the two majority species in the resulting mixture) is approximately ~16%.

It should be possible to improve this yield by tweaking the synthesis design to keep the catalyst in better activation and reduction conditions. This might be done either (i) by decreasing the flow rate feeding the exchange reactor; or (ii) by finding strategies to more effectively prevent the formation of carbon filaments on the catalyst. It is in these amorphous carbon filaments where most of the $CH_4$ introduced into the system in the first step of the synthesis is lost (remember that these catalysts are used precisely for methane cracking). Some of it remains on the catalyst, and is removed each time the catalyst is regenerated between synthesis steps. Some of the filament fragments are observed in the mass spectra as the families of $C_3Q_{8-2z}$, $C_4Q_{10-2z}$, $C_5Q_{12-2z}$, etc.

Both the amount of mixture (~5-6×10$^3$ cm$^3$) and the concentration of $CT_xH_{(4-x)}$ are significantly higher than those utilized in previous studies. This gives us the opportunity to record Raman spectra with further increased resolution, in order to isolate individual $Q_1$-branches of the symmetric normal mode of vibration, $\nu_1$, of each of the species present in the mixture. With this one should be able to estimate the amount of each of them more rigorously.

## V. CONCLUSION

Our results have shown that $CQ_4$ can be produced in large enough quantities so that, from an individual production batch. Therefore, a wealth of experiments can be performed using the same starting sample. Such experiments may address aspects of different analysis techniques (as demonstrated here); or the gas may be used for specific chemical interaction scenarios, such as for example exposure of particular surfaces to tritiated methane. But this work has also shown, that our approach – while rather successful – is still not yet perfect.

Since the production of $CQ_4$ via the CAPER facility had to compete with its priority use for the KATRIN experiment, the time available for incremental improvements of $CQ_4$ processing procedures was rather limited. However, with the experience gained thus far we envisage to coordinate our future efforts better with the KATRIN service periods, during which CAPER can be utilized for different tasks. In this way we expect to gain even better control over the distribution of $CQ_4$-isotolologues in the mixture as was possible to date. Even better for the production of "pure" CT4 – as would be ideal – would be to use a H-free production path. This, however, would require the use of a different chemical reactor setup; this would be a long-term goal since at present the perpetual use of CAPER for other experiments prevents us from implementing structural changes.

Equally, our Raman analysis warrants attention as well; this is because the TRIHYDE Raman analysis tool was designed to analyze $Q_2$-mixtures, rather than performing high-resolution spectroscopy. Thus, in particular it is planned to add high-resolution Raman spectrometer which would allow for full spectral resolution of all $CT_xH_{(4-x)}$ species; at present, parts of the spectral features do overlap.

In this context, we have just begun theoretical quantum calculations on the structure of the Raman vibration-rotation spectra of each species in the $CT_xH_{(4-x)}$ family. The aim is to be able to subtract these theoretical calculations from the experimental spectra, in order both to reduce the non-$Q_1$ branch background, and to better attribute the contribution of each species in the family to the overall Raman spectrum (unfortunately, individual $S_1$-, $R_1$-, $P_1$- and $O_1$- branches overlap and cannot be easily separated). With the help of the "pure" synthetic Raman spectra, in comparison to the "mixed" experimental Raman spectra, we then hope to be able to disentangle and quantify all individual $CT_xH_{(4-x)}$ species in our gas mixtures.

**Disclosure Statement**

The authors reported no potential competing interest.

**Acknowledgements**

We gratefully acknowledge the financial support for research visits by D. Diaz Barrero to the TLK, provided by the KIT Center for Elementary Particle and Astroparticle Physics (KCETA).